\documentclass[aps, pre, a4paper,twocolumn, superscriptaddress]{revtex4}

\usepackage{float}
\usepackage{graphicx}

\newcommand{\msd}{{\cal S}(t)}
\newcommand{\td}{t_D}

\newcommand {\be} {\begin{equation}}
\newcommand {\ee} {\end{equation}}
\newcommand {\bea} {\begin{eqnarray}}
\newcommand {\eea} {\end{eqnarray}}

\begin{document}

\title{Single-file diffusion with non-thermal initial conditions}

\author{Ludvig Lizana}
\affiliation{Integrated Science Lab, Department of Physics, Ume{\aa}
University, SE-901 87 Ume{\aa}, Sweden}
 
\author{Michael A. Lomholt}
\affiliation{MEMPHYS - Center for Biomembrane Physics, Department of Physics, Chemistry and Pharmacy,
University of Southern Denmark, Campusvej 55, 5230 Odense M, Denmark}

\author{Tobias Ambj\"ornsson}
\affiliation{Department of Astronomy and Theoretical Physics, Lund University,
  S\"olvegatan 14A, SE-223 62 Lund, Sweden, \& Department of Chemistry,
  Massachusetts Institute of Technology, 77 Massachusetts Avenue, Cambridge,
  MA 02139, USA}

\date{\today}

%

\begin{abstract}
  Single-file diffusion is a theoretically challenging many-body problem where
  the calculation of even the simplest observables, e.g. mean square
  displacement, for a tracer particle requires a heavy mathematical
  machinery. There is therefore a need for simple approaches which predict
  qualitatively correct behaviours. Here we put forward one such method which
  we use to investigate the influence of non-thermal initial conditions on the
  dynamics of a tracer particle. With our new approach we reproduce, up to
  scaling, several known asymptotic results for the tracer particle mean
  square displacement.
\end{abstract}

\maketitle
%
%

\section{Introduction}

The dynamics of diffusing particles in one dimension which are unable to pass
each other, often referred to as single-file diffusion, has attracted great
theoretical interest for at least 50 years, see \cite{harris1965diffusion,
  levitt1973dynamics, kehr1981diffusion, percus1974anomalous,
  hahn1999propagator, van1985mean, schutz1997exact, burlatsky1996motion,
  alexander1978diffusion, jara2006nonequilibrium, arratia1983motion,
  kollmann2003single, lizana2008single, taloni2008langevin}. More recent
progress in the field addresses single-file diffusion in a potential
\cite{barkai2009theory} and that of particles with different diffusion
constants \cite{lomholt2011dissimilar}.  One of the main theoretical
predictions is that a tagged (or a tracer) particle, explores the system
subdiffusively even though the collective behaviour is identical to that of
non-interacting particles \cite{rodenbeck1998calculating,
  lomholt2011dissimilar}. The subdiffusive behaviour can be understood through the tracer particle's
velocity-velocity correlation function which is negative and decays as $ \td^{-3/2}$
with delay time $\td$ \cite{percus1974anomalous}; this results translates into the classical  square root law 
for the tagged particle's mean square discplacement $\msd\propto t^{1/2}$ where  $t$ is time.

Nowadays single-file diffusion also finds experimental realizations. A few
examples are diffusion  in colloidal systems
\cite{wei2000single,lin2005random}, transport in microporous materials
\cite{kukla1996nmr, meersmann2000exploring}, and permeability of potassium
ions in nerve fibres \cite{hodgkin1955potassium}. Also, exclusion effects have
been proposed to be of importance for transcription factor (DNA binding
proteins) dynamics \cite{li2009effects}.

In order to calculate dynamical properties of a tagged particle in a
single-file system one needs to deal with a theoretically challenging
many-body problem. This typically requires rather elaborate mathematics
(e.g. \cite{harris1965diffusion, levitt1973dynamics, kollmann2003single,
  lizana2008single, barkai2009theory}) which may cloud fundamental
understanding. There is therefore a need for simplified methods that captures
correct qualitative behaviour and at the same time provide new insights to the
problem.  A few such papers already exist \cite{alexander1978diffusion,
  taloni2008langevin} for thermal initial conditions and identical
particles.
To the best of our knowledge there has not been a similar progress on simple approaches for non-thermal initial conditions. 
Some aspects have, however, been addressed \cite{barkai2010diffusion, flomenbom2008single, lizana2010foundation} using rather involved techniques. 
The main goal of this paper is to put forward a new but simple method to
calculate tagged particle properties, in particular  $\msd$, under non-thermal conditions. Our approach relies on the assumption that we can make a quasi-static approximation for the velocity-velocity autocorrelation function and set it equal to the correlation function at equilibrium.
Based on this we are able to reproduce known scaling behaviours pertaining to
different types of initial conditions.

%
%
\section{The quasi-equilibrium approach}

Imagine a single-file system extending to infinity in both directions. The
particles are identical, point-like and undergo Brownian motion in between
hard-core repulsive interactions. Now we tag one particle, label it ``0'', and
place it at time $t=0$ in the origin, that is, $x_0(t=0)=0$. The remaining
particles are initially distributed symmetrically around $x_0=0$,
$...x_{-2},x_{-1}, x_0,x_1,x_2,...$, according to some non-thermal
distribution $\rho_0(x)$.  We then ask: what is the mean square displacement,
$\msd $, of particle $0$? Here we define $\msd = \langle x_0(t)^2\rangle$ where $\langle...\rangle$ denotes ensemble average.

Our expression for $\msd$ is based on the following quasi-equilibrium argument.  For an
equilibrated single-file it is well known that the long time limit of the velocity-velocity correlation function for a tagged particle
 is \cite{percus1974anomalous}
\be\label{eq:vvthermal}
\langle v(t) v(t-\td) \rangle  \simeq -\sqrt{\frac{D}{\pi}} \frac{1}{4\varrho} |\td|^{-3/2}
\ee
where $v(t)=dx(t)/dt$ and $\varrho$ is the constant average particle
concentration (average inverse inter-particle distance), and $D$ is the
single-particle diffusion constant (assumed equal for all particles).
Now we assume that Eq. (\ref{eq:vvthermal}) holds for all, even non-thermal, initial conditions if we replace $\varrho$ with the time-dependent concentration
$\rho(x,t)$ in the vicinity of the tagged particle, i.e., 
\be\label{eq:quasistaticapprox}
\varrho \rightarrow \rho(x=0,t)
\ee
This is a quasi-equilibrium approach which we assume to hold since concentration relaxations propagate faster than the tracer particle
\footnote{The average distance covered by a tagged particle scales as $(Dt)^{1/4}$ whereas the particle density close to equilibrium is
$\rho (x,t)\simeq \rho (x,t=\infty) [1+ {\rm const.}/\sqrt{Dt}]$. Tracer particle dynamics is therefore much slower than concentration  relaxations. See also \cite{lizana2010foundation} for more details.}.
Integrating Eq. (\ref{eq:vvthermal}) twice with respect to time yields 
$\msd = \int_0^t dt' \int_0^t dt''\langle v(t')v(t'')\rangle = 2 \int_0^t dt' \int_0^{t'} du \langle v(t')v(t'-u)\rangle$. If we furthermore assume the tagged particle behaves elastically over time-scales longer than the velocity relaxation,  we have $\int_0^\infty du \langle v(t) v(t-u) \rangle=0$
\footnote{This is equivalent to the mobility vanishing at low frequencies, which holds for the single-file system in equilibrium, see for instance \cite{taloni2008langevin}.}. 
This means that we can write $\msd = -2 \int_0^t dt' \int_{t'}^\infty du \langle v(t')v(t'-u)\rangle$, which together with Eq. (\ref{eq:vvthermal}) and $\varrho=\rho(x=0,t')$ gives
\be\label{eq:msd} 
\msd \simeq \sqrt{\frac{D}{\pi}} \int_0^t dt'
\frac{1}{\sqrt{t'}\rho(x=0,t')} 
\ee 
Now, since the macroscopic particle concentration in a single-file system and
in a system of non-interacting particles are indistinguishable
\cite{rodenbeck1998calculating} we can calculate $\rho(x= 0,t)$ by simply
solving the diffusion equation in one dimension, and subsequently setting
$x=0$ in the associated solution. In effect this amounts to convoluting
$\rho_0(x_0)$ with a Gaussian propagator
\footnote{Consider diffusion on an
  infinite line with the initial density $\rho_0(x)$. Here the solution to the
  diffusion equation, $\partial \rho(x,t)/\partial t = D \partial^2
  \rho(x,t)/\partial x^2$, gives a solution for density according to
  $\rho(x,t) = \int_{-\infty}^\infty \frac{e^{-(x-y)^2/(4Dt)}}{\sqrt{4\pi D
      t}} \rho_0(y)\, dy$}
\be\label{eq:rho_x0}
\rho(x=0,t) = \int_{-\infty}^\infty \frac{e^{-x_0^2/(4Dt)}}{\sqrt{4\pi D t}}
\rho_0(x_0)\, dx_0,
\ee
Equation (\ref{eq:msd}) [together with Eq. (\ref{eq:rho_x0})] constitutes our main result. It is
straightforward to see that for uniform initial densities those equations lead
to classical single-file result for the mean square displacement
\be\label{eq:msd_harris}
\msd \simeq \frac{1}{\varrho} \sqrt{\frac{4Dt} \pi} 
\ee
(see e.g. \cite{harris1965diffusion}) as it should.

\section{Special cases}

In this section we will investigate four different types of  $\rho_0 (x_0)  $
and show that our simple model gives  correct results up to scaling in
$t$. Our results are summarised briefly in Tab. \ref{table1} and depicted in
Fig. \ref{figure1}.


\subsection{Exponential distribution }

Here we consider the case where the initial particle density decays exponentially from the origin 
\be
\rho_0(x_0) = \frac{N} {2\Delta} e^{-|x_0|/\Delta},
\ee
where $N$ is the number of particles and $\Delta$ is the characteristic decay length. Using this expression in Eq. (\ref{eq:rho_x0}) leads to
\be\label{eq:rhoexp}
\rho(x=0,t) = \frac{N}{2\Delta} e^{t/\tau} {\rm erfc} \left(\sqrt{t/\tau} \right) 
\ee
which together with Eq. (\ref{eq:msd}) gives $\msd$ predicted by our model. Here we
introduced 
\be\label{eq:tau}
\tau = \frac{\Delta^2}{D}
\ee
Figure \ref{figure1} (solid blue line) shows the result of a numerical integration of Eq. (\ref{eq:msd}) [using Eq. (\ref{eq:rhoexp})]. We see clearly that $\msd$ exhibits two regions with different dynamics with crossover time $\tau$.
The limiting behaviours can be found analytically if we expand
Eq. (\ref{eq:rhoexp}) in the short and long time limit: for short times ($t\ll
\tau$) we get $\rho(x=0,t) \simeq N/(2\Delta)$, whereas for long times ($t\gg
\tau$) $\rho(x=0,t) \simeq N/\sqrt{4\pi Dt}$. The resulting mean square
displacement in the two limits reads
\be
\msd \simeq \Bigg\{ 
\begin{array} {lll} 
	\frac{2\Delta}{N}\sqrt{\frac{4Dt}\pi} &\propto \sqrt t,  &\qquad t \ll \tau \\ 
	2Dt/N &\propto t ,           &\qquad t\gg \tau 
\end{array}
\ee
This means that for short times the system is crowded enough such that we
recover the known scaling with time in Eq. (\ref{eq:msd_harris}).  For long
times the tagged particle follows the system's center-of-mass motion yielding
$\msd \sim Dt/N$. For this limit we point out that C. Aslangul
\cite{aslangul2007classical} obtained $\pi Dt/N$ using a more elaborate
approach.
\begin{table}
\begin{tabular}{|c|c|c|}
\hline
$\rho_0(x_0)$  &\multicolumn{2}{c|} {$\msd\propto $ } \\
 & $t/\tau\ll1$ & $t/\tau\gg1$\\
\hline
$\frac{N}{2 \Delta} e^{-|x_0|/\Delta}$  & $ \sqrt{t}$  & $  t$\\
&  &\\
$\frac{N}{\sqrt{2 \pi \Delta^2} }e^{-x_0^2/(2\Delta^2)}$ & $ \sqrt {t}$  &
$ t$ \\
&  &\\
$\sum_{n=-\infty}^\infty\delta(x_0-n\Delta)$ & $ t$& $  \sqrt{t}$ \\
&\multicolumn{2}{c|} { }\\
$\frac 1 \Delta \left(\frac{|x_0|}{\Delta}\right)^{-\beta}$ &
\multicolumn{2}{c|} {$ t^{(1+\beta)/2}$ } \\
  &\multicolumn{2}{c|} { } \\
\hline
\end{tabular}
\caption{Summary of results ($\tau = \Delta^2/D$).}
\label{table1}
\end{table}

\subsection{Gaussian distribution }

In this subsection we take the particles to be initially distributed as a
Gaussian around the origin
\be
\rho_0(x_0) = \frac{N}{\sqrt{2 \pi \Delta^2} }e^{-x_0^2/(2\Delta^2)}.
\ee
Putting this expression  into Eq. (\ref{eq:rho_x0}) gives
\be
\rho(x=0,t) = \frac N \Delta \sqrt{ \frac 1 {2\pi (2t/\tau+1)} }
\ee
with $\tau$ given by Eq. (\ref{eq:tau}). The resulting $\msd$ from Eq. (\ref{eq:msd}) is shown in Fig. \ref{figure1} (solid red line) where we see, again, two dynamically different regions with crossover time $\tau$. A similar analysis as in the previous subsection shows that
\be
\msd \simeq \Bigg\{ 
\begin{array} {lll} 
	\Delta\sqrt{8Dt}/N &\propto \sqrt t,  & \qquad t \ll \tau \\ 
	2Dt/N &\propto t,           & \qquad t \gg \tau 
\end{array}
\ee
where we used that  $\rho(x=0,t) \simeq N/(\Delta \sqrt{2\pi})$ for short times and $\rho(x=0,t) \simeq N/\sqrt{4\pi Dt}$ for long times.  The crossover time as well as the limiting behaviours of $\msd$ are up to scaling in agreement with \cite{barkai2010diffusion}.

\subsection{Equidistant particles}

\begin{figure}
\includegraphics[width=\columnwidth]{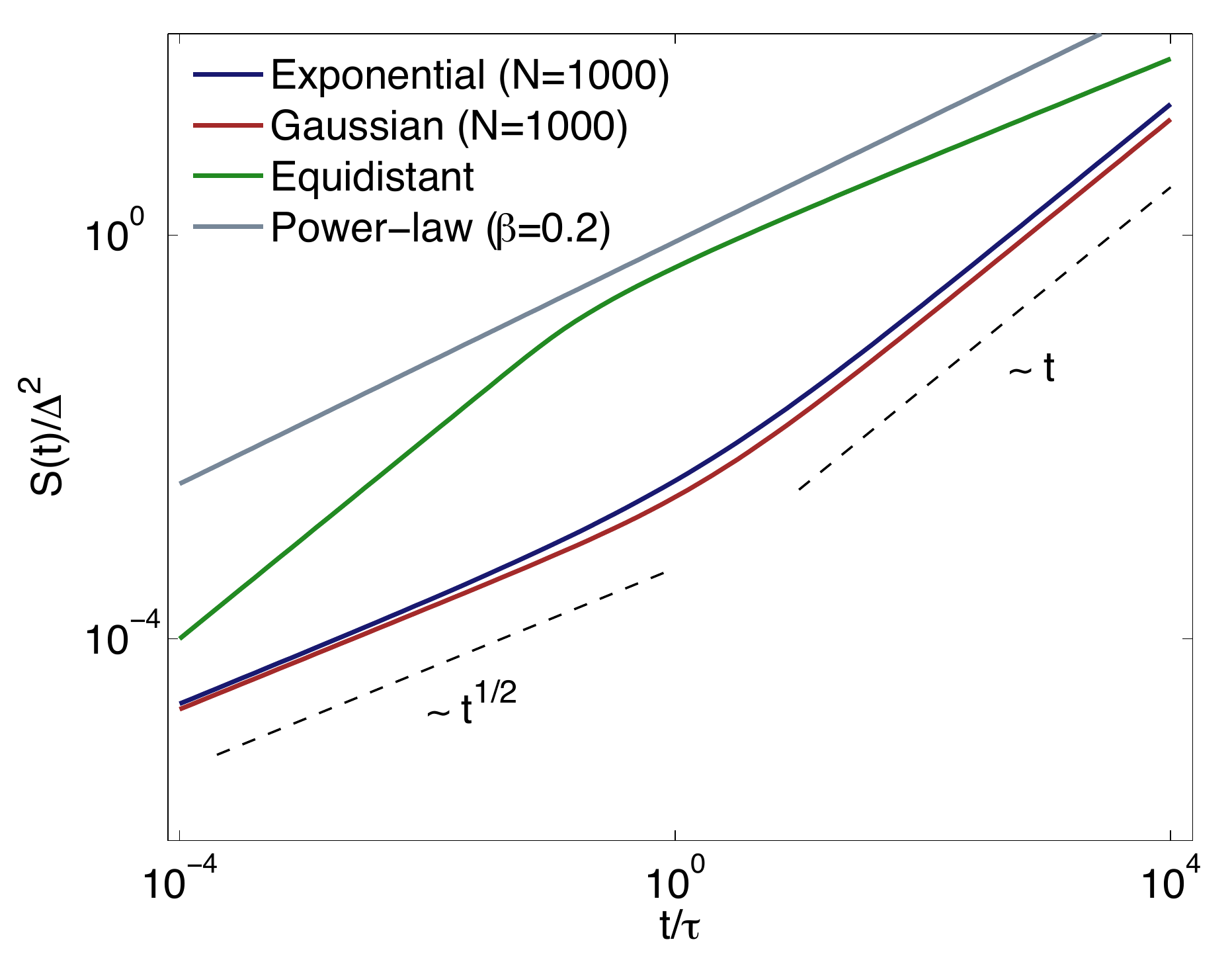}
\caption{Mean square displacement versus time for different initial distributions $\rho_0(x_0)$ ($\tau=\Delta^2/D$). The curves are obtained from a numerical integration of Eq. (\ref{eq:msd})}
\label{figure1}
\end{figure}

Here we address the situation where the particles are initially distributed on
the line with the same distance $\Delta$ apart from each other:
\be
\rho_0(x_0) = \sum_{n=-\infty}^\infty\delta(x_0-n\Delta)
\ee
where $\delta(z)$ denotes the Dirac delta function. This case is different in
one important aspect compared to the other cases in this paper: the
concentration $\rho(x,t)$ in the long time limit approaches the (finite)
constant $1/\Delta$ whereas in the other cases considered here $\rho(x,t)$
goes to zero. This means that we expect $\msd \sim \sqrt {Dt}$ for long times
for the present case. If we carry out the calculation of $\rho(x=0,t)$ as
before we find
\be
\rho(x=0,t) = \frac{1}{\Delta} \frac{\vartheta_3 (\sqrt{4t/\tau})} { \sqrt{4\pi t/\tau} }
\ee
where $\vartheta_3(z)$ is the elliptic Jacobi-Theta function \cite{abramowitz1965handbook}
\be
\vartheta_3(z)= 1+2\sum_{n=1}^\infty e^{-(n/z)^2}
\ee
The behaviour of $\msd$ is depicted in Fig. \ref{figure1} (solid green line)
where two regimes are visible once more. Using that $\rho(x=0,t) \simeq
1/\sqrt{4\pi D t}$ for short times and $\rho(x=0,t) \simeq 1/\Delta$ in the
long time limit, we obtain the short and long time asymptotics:
\be
\msd \simeq \Bigg\{ 
\begin{array} {lll} 
	2Dt &\propto t,  			&\qquad t\ll \tau \\ 
	\Delta \sqrt{\frac{4Dt} \pi} &\propto \sqrt t, &\qquad t \gg \tau 
\end{array}
\ee
The long time limit is in agreement with \cite{lizana2010foundation} up to scaling. However, the prefactor is a factor $\sqrt 2$ too large compared to the correct result. Our simple model is therefore unable to account for numerical factors but gives the correct scaling behaviour in all cases we examined here. 

One reason to why our method is unable to produce proper prefactors is that the density $\rho(x,t)$ does not contain all necessary information to predict the motion of the tagged particle. One way of seeing this is from the following simple example. Imagine that all particles are placed equidistantly with inter-particle distance $\Delta$. Now we draw a random number between zero and $\Delta$ from a uniform distribution and add the same random number to the position of all particles. If we average over many such subsystems we would obtain a new system with uniform density $1/\Delta$ just as in thermal equilibrium (each single-equidistant case has of course  Dirac delta-peaked density). However, since each system is just the equidistant case shifted by a constant, $\msd$ for the averaged system will still be a factor $\sqrt{2}$ different compared to the case of thermal initial conditions. This illustrates that two cases may have the same uniform density but yet yield different prefactors to the $\sqrt t$ behaviour. Thus, the density $\rho(x,t)$ is by itself not sufficient to predict exact numerical prefactors. 
Another way of seeing that $\rho(x,t)$ does not hold all necessary information is
to consider $\msd$ in thermal equilibrium. For this case Alexander \& Pincus \cite{alexander1978diffusion} argued that $\msd$ can be related to the density-density correlation function (dynamic structure factor) rather than the density itself. Their relation is another indication that only the density is not sufficient to predict the full $\msd$ but that one also need to consider density fluctuations at different locations.

\subsection{Power-law distribution }

Here we investigate the scale-free initial density
\be\label{eq:rhoPL}
\rho_0(x_0) = \frac 1 \Delta \left(\frac{|x_0|}{\Delta}\right)^{-\beta}, \ \xi < |x_0| < \infty, \ 0\leq \beta \leq1
\ee
where we take the cut-off $\xi\to 0$. The power-law form (\ref{eq:rhoPL}) yields standard single-file dynamics [Eq. (\ref{eq:msd_harris})] for $\beta=0$. If we put Eq. (\ref{eq:rhoPL}) into Eq. (\ref{eq:rho_x0}) we find
\be
\rho(x=0,t) = \frac{1}{\sqrt{\pi}\Delta}  \Gamma \left(\frac{1-\beta}2 \right) 
\left(\sqrt{4t/\tau}\right)^{-\beta}
\ee
where $\Gamma (z)$ is the Gamma function. Using this formula in
Eq. (\ref{eq:msd})  leads to a mean
square displacement characterized by an exponent in between that of standard
and single-file diffusion. Explicitly
\be
\msd \simeq \Delta^2  \frac{2^{1+\beta}}{(1+\beta)\Gamma \left(\frac{1-\beta}2 \right)} (t/\tau)^{(\beta+1)/2}
\propto t^{(\beta+1)/2}
\ee
which agrees with \cite{barkai2010diffusion,flomenbom2008single} up to scaling.


%
%
\section{Concluding remarks}

In this paper we have analysed the effect of non-thermal initial conditions on
the dynamics of a tagged particle in a single-file diffusion system. We used a
new simple method based on a quasi-equilibrium assumption and predicted the
scaling behaviour and crossover times known from literature. A future
challenge will be to try to improve our method to also yield correct
prefactors.  The reason to why the approach works so well is that particle
density relaxations decays faster than the motion of a tracer particle. It
would be interesting to see how this assumption holds in other settings such a
disordered single file where the diffusion constants are assigned randomly.

\section{Acknowledgements}
We thank Ophir Flomenbom for helpful discussions. LL acknowledges the Knut and
Alice Wallenberg (KAW) foundation for financial support. TA is grateful to KAW
and the Swedish Research Council for funding.


\begin{thebibliography}{28}
\expandafter\ifx\csname natexlab\endcsname\relax\def\natexlab#1{#1}\fi
\expandafter\ifx\csname bibnamefont\endcsname\relax
  \def\bibnamefont#1{#1}\fi
\expandafter\ifx\csname bibfnamefont\endcsname\relax
  \def\bibfnamefont#1{#1}\fi
\expandafter\ifx\csname citenamefont\endcsname\relax
  \def\citenamefont#1{#1}\fi
\expandafter\ifx\csname url\endcsname\relax
  \def\url#1{\texttt{#1}}\fi
\expandafter\ifx\csname urlprefix\endcsname\relax\def\urlprefix{URL }\fi
\providecommand{\bibinfo}[2]{#2}
\providecommand{\eprint}[2][]{\url{#2}}

\bibitem[{\citenamefont{Harris}(1965)}]{harris1965diffusion}
\bibinfo{author}{\bibfnamefont{T.}~\bibnamefont{Harris}},
  \bibinfo{journal}{Journal of Applied Probability}
  \textbf{\bibinfo{volume}{2}}, \bibinfo{pages}{323} (\bibinfo{year}{1965}).

\bibitem[{\citenamefont{Levitt}(1973)}]{levitt1973dynamics}
\bibinfo{author}{\bibfnamefont{D.}~\bibnamefont{Levitt}},
  \bibinfo{journal}{Physical Review A} \textbf{\bibinfo{volume}{8}},
  \bibinfo{pages}{3050} (\bibinfo{year}{1973}).

\bibitem[{\citenamefont{Kehr et~al.}(1981)\citenamefont{Kehr, Kutner, and
  Binder}}]{kehr1981diffusion}
\bibinfo{author}{\bibfnamefont{K.}~\bibnamefont{Kehr}},
  \bibinfo{author}{\bibfnamefont{R.}~\bibnamefont{Kutner}}, \bibnamefont{and}
  \bibinfo{author}{\bibfnamefont{K.}~\bibnamefont{Binder}},
  \bibinfo{journal}{Physical Review B} \textbf{\bibinfo{volume}{23}},
  \bibinfo{pages}{4931} (\bibinfo{year}{1981}).

\bibitem[{\citenamefont{Percus}(1974)}]{percus1974anomalous}
\bibinfo{author}{\bibfnamefont{J.}~\bibnamefont{Percus}},
  \bibinfo{journal}{Physical Review A} \textbf{\bibinfo{volume}{9}},
  \bibinfo{pages}{557} (\bibinfo{year}{1974}).

\bibitem[{\citenamefont{Hahn and Karger}(1999)}]{hahn1999propagator}
\bibinfo{author}{\bibfnamefont{K.}~\bibnamefont{Hahn}} \bibnamefont{and}
  \bibinfo{author}{\bibfnamefont{J.}~\bibnamefont{Karger}},
  \bibinfo{journal}{Journal of Physics A: Mathematical and General}
  \textbf{\bibinfo{volume}{28}}, \bibinfo{pages}{3061} (\bibinfo{year}{1999}).

\bibitem[{\citenamefont{van Beijeren and Kutner}(1985)}]{van1985mean}
\bibinfo{author}{\bibfnamefont{H.}~\bibnamefont{van Beijeren}}
  \bibnamefont{and} \bibinfo{author}{\bibfnamefont{R.}~\bibnamefont{Kutner}},
  \bibinfo{journal}{Physical review letters} \textbf{\bibinfo{volume}{55}},
  \bibinfo{pages}{238} (\bibinfo{year}{1985}).

\bibitem[{\citenamefont{Sch{\"u}tz}(1997)}]{schutz1997exact}
\bibinfo{author}{\bibfnamefont{G.}~\bibnamefont{Sch{\"u}tz}},
  \bibinfo{journal}{Journal of statistical physics}
  \textbf{\bibinfo{volume}{88}}, \bibinfo{pages}{427} (\bibinfo{year}{1997}).

\bibitem[{\citenamefont{Burlatsky et~al.}(1996)\citenamefont{Burlatsky,
  Oshanin, Moreau, and Reinhardt}}]{burlatsky1996motion}
\bibinfo{author}{\bibfnamefont{S.}~\bibnamefont{Burlatsky}},
  \bibinfo{author}{\bibfnamefont{G.}~\bibnamefont{Oshanin}},
  \bibinfo{author}{\bibfnamefont{M.}~\bibnamefont{Moreau}}, \bibnamefont{and}
  \bibinfo{author}{\bibfnamefont{W.}~\bibnamefont{Reinhardt}},
  \bibinfo{journal}{Physical Review E} \textbf{\bibinfo{volume}{54}},
  \bibinfo{pages}{3165} (\bibinfo{year}{1996}).

\bibitem[{\citenamefont{Alexander and Pincus}(1978)}]{alexander1978diffusion}
\bibinfo{author}{\bibfnamefont{S.}~\bibnamefont{Alexander}} \bibnamefont{and}
  \bibinfo{author}{\bibfnamefont{P.}~\bibnamefont{Pincus}},
  \bibinfo{journal}{Physical Review B} \textbf{\bibinfo{volume}{18}},
  \bibinfo{pages}{2011} (\bibinfo{year}{1978}).

\bibitem[{\citenamefont{Jara and Landim}(2006)}]{jara2006nonequilibrium}
\bibinfo{author}{\bibfnamefont{M.}~\bibnamefont{Jara}} \bibnamefont{and}
  \bibinfo{author}{\bibfnamefont{C.}~\bibnamefont{Landim}}, in
  \emph{\bibinfo{booktitle}{Annales de l'Institut Henri Poincare (B)
  Probability and Statistics}} (\bibinfo{organization}{Elsevier},
  \bibinfo{year}{2006}), vol.~\bibinfo{volume}{42}, pp.
  \bibinfo{pages}{567--577}.

\bibitem[{\citenamefont{Arratia}(1983)}]{arratia1983motion}
\bibinfo{author}{\bibfnamefont{R.}~\bibnamefont{Arratia}},
  \bibinfo{journal}{The Annals of Probability} pp. \bibinfo{pages}{362--373}
  (\bibinfo{year}{1983}).

\bibitem[{\citenamefont{Kollmann}(2003)}]{kollmann2003single}
\bibinfo{author}{\bibfnamefont{M.}~\bibnamefont{Kollmann}},
  \bibinfo{journal}{Physical review letters} \textbf{\bibinfo{volume}{90}},
  \bibinfo{pages}{180602} (\bibinfo{year}{2003}).

\bibitem[{\citenamefont{Lizana and Ambj{\"o}rnsson}(2008)}]{lizana2008single}
\bibinfo{author}{\bibfnamefont{L.}~\bibnamefont{Lizana}} \bibnamefont{and}
  \bibinfo{author}{\bibfnamefont{T.}~\bibnamefont{Ambj{\"o}rnsson}},
  \bibinfo{journal}{Physical review letters} \textbf{\bibinfo{volume}{100}},
  \bibinfo{pages}{200601} (\bibinfo{year}{2008}).

\bibitem[{\citenamefont{Taloni and Lomholt}(2008)}]{taloni2008langevin}
\bibinfo{author}{\bibfnamefont{A.}~\bibnamefont{Taloni}} \bibnamefont{and}
  \bibinfo{author}{\bibfnamefont{M.~A.} \bibnamefont{Lomholt}},
  \bibinfo{journal}{Physical Review E} \textbf{\bibinfo{volume}{78}},
  \bibinfo{pages}{051116} (\bibinfo{year}{2008}).

\bibitem[{\citenamefont{Barkai and Silbey}(2009)}]{barkai2009theory}
\bibinfo{author}{\bibfnamefont{E.}~\bibnamefont{Barkai}} \bibnamefont{and}
  \bibinfo{author}{\bibfnamefont{R.}~\bibnamefont{Silbey}},
  \bibinfo{journal}{Physical review letters} \textbf{\bibinfo{volume}{102}},
  \bibinfo{pages}{50602} (\bibinfo{year}{2009}).

\bibitem[{\citenamefont{Lomholt et~al.}(2011)\citenamefont{Lomholt, Lizana, and
  Ambj{\"o}rnsson}}]{lomholt2011dissimilar}
\bibinfo{author}{\bibfnamefont{M.}~\bibnamefont{Lomholt}},
  \bibinfo{author}{\bibfnamefont{L.}~\bibnamefont{Lizana}}, \bibnamefont{and}
  \bibinfo{author}{\bibfnamefont{T.}~\bibnamefont{Ambj{\"o}rnsson}},
  \bibinfo{journal}{The Journal of chemical physics}
  \textbf{\bibinfo{volume}{134}}, \bibinfo{pages}{045101}
  (\bibinfo{year}{2011}).

\bibitem[{\citenamefont{R{\"o}denbeck et~al.}(1998)\citenamefont{R{\"o}denbeck,
  K{\"a}rger, and Hahn}}]{rodenbeck1998calculating}
\bibinfo{author}{\bibfnamefont{C.}~\bibnamefont{R{\"o}denbeck}},
  \bibinfo{author}{\bibfnamefont{J.}~\bibnamefont{K{\"a}rger}},
  \bibnamefont{and} \bibinfo{author}{\bibfnamefont{K.}~\bibnamefont{Hahn}},
  \bibinfo{journal}{Physical Review E} \textbf{\bibinfo{volume}{57}},
  \bibinfo{pages}{4382} (\bibinfo{year}{1998}).

\bibitem[{\citenamefont{Wei et~al.}(2000)\citenamefont{Wei, Bechinger, and
  Leiderer}}]{wei2000single}
\bibinfo{author}{\bibfnamefont{Q.}~\bibnamefont{Wei}},
  \bibinfo{author}{\bibfnamefont{C.}~\bibnamefont{Bechinger}},
  \bibnamefont{and} \bibinfo{author}{\bibfnamefont{P.}~\bibnamefont{Leiderer}},
  \bibinfo{journal}{Science } \textbf{\bibinfo{volume}{287}},
  \bibinfo{pages}{625} (\bibinfo{year}{2000}).

\bibitem[{\citenamefont{Lin et~al.}(2005)\citenamefont{Lin, Meron, Cui, Rice,
  and Diamant}}]{lin2005random}
\bibinfo{author}{\bibfnamefont{B.}~\bibnamefont{Lin}},
  \bibinfo{author}{\bibfnamefont{M.}~\bibnamefont{Meron}},
  \bibinfo{author}{\bibfnamefont{B.}~\bibnamefont{Cui}},
  \bibinfo{author}{\bibfnamefont{S.}~\bibnamefont{Rice}}, \bibnamefont{and}
  \bibinfo{author}{\bibfnamefont{H.}~\bibnamefont{Diamant}},
  \bibinfo{journal}{Physical review letters} \textbf{\bibinfo{volume}{94}},
  \bibinfo{pages}{216001} (\bibinfo{year}{2005}).

\bibitem[{\citenamefont{Kukla et~al.}(1996)\citenamefont{Kukla, Kornatowski,
  Demuth, Girnus, Pfeifer, Rees, Schunk, Unger, Karger et~al.}}]{kukla1996nmr}
\bibinfo{author}{\bibfnamefont{V.}~\bibnamefont{Kukla}},
  \bibinfo{author}{\bibfnamefont{J.}~\bibnamefont{Kornatowski}},
  \bibinfo{author}{\bibfnamefont{D.}~\bibnamefont{Demuth}},
  \bibinfo{author}{\bibfnamefont{I.}~\bibnamefont{Girnus}},
  \bibinfo{author}{\bibfnamefont{H.}~\bibnamefont{Pfeifer}},
  \bibinfo{author}{\bibfnamefont{L.}~\bibnamefont{Rees}},
  \bibinfo{author}{\bibfnamefont{S.}~\bibnamefont{Schunk}},
  \bibinfo{author}{\bibfnamefont{K.}~\bibnamefont{Unger}},
  \bibinfo{author}{\bibfnamefont{J.}~\bibnamefont{Karger}},
  \bibnamefont{et~al.}, \bibinfo{journal}{Science (New York, NY)}
  \textbf{\bibinfo{volume}{272}}, \bibinfo{pages}{702} (\bibinfo{year}{1996}).

\bibitem[{\citenamefont{Meersmann et~al.}(2000)\citenamefont{Meersmann, Logan,
  Simonutti, Caldarelli, Comotti, Sozzani, Kaiser, and
  Pines}}]{meersmann2000exploring}
\bibinfo{author}{\bibfnamefont{T.}~\bibnamefont{Meersmann}},
  \bibinfo{author}{\bibfnamefont{J.}~\bibnamefont{Logan}},
  \bibinfo{author}{\bibfnamefont{R.}~\bibnamefont{Simonutti}},
  \bibinfo{author}{\bibfnamefont{S.}~\bibnamefont{Caldarelli}},
  \bibinfo{author}{\bibfnamefont{A.}~\bibnamefont{Comotti}},
  \bibinfo{author}{\bibfnamefont{P.}~\bibnamefont{Sozzani}},
  \bibinfo{author}{\bibfnamefont{L.}~\bibnamefont{Kaiser}}, \bibnamefont{and}
  \bibinfo{author}{\bibfnamefont{A.}~\bibnamefont{Pines}},
  \bibinfo{journal}{The Journal of Physical Chemistry A}
  \textbf{\bibinfo{volume}{104}}, \bibinfo{pages}{11665}
  (\bibinfo{year}{2000}).

\bibitem[{\citenamefont{Hodgkin and Keynes}(1955)}]{hodgkin1955potassium}
\bibinfo{author}{\bibfnamefont{A.}~\bibnamefont{Hodgkin}} \bibnamefont{and}
  \bibinfo{author}{\bibfnamefont{R.}~\bibnamefont{Keynes}},
  \bibinfo{journal}{The Journal of physiology} \textbf{\bibinfo{volume}{128}},
  \bibinfo{pages}{61} (\bibinfo{year}{1955}).

\bibitem[{\citenamefont{Li et~al.}(2009)\citenamefont{Li, Berg, and
  Elf}}]{li2009effects}
\bibinfo{author}{\bibfnamefont{G.}~\bibnamefont{Li}},
  \bibinfo{author}{\bibfnamefont{O.}~\bibnamefont{Berg}}, \bibnamefont{and}
  \bibinfo{author}{\bibfnamefont{J.}~\bibnamefont{Elf}},
  \bibinfo{journal}{Nature Physics} \textbf{\bibinfo{volume}{5}},
  \bibinfo{pages}{294} (\bibinfo{year}{2009}).

\bibitem[{\citenamefont{Barkai and Silbey}(2010)}]{barkai2010diffusion}
\bibinfo{author}{\bibfnamefont{E.}~\bibnamefont{Barkai}} \bibnamefont{and}
  \bibinfo{author}{\bibfnamefont{R.}~\bibnamefont{Silbey}},
  \bibinfo{journal}{Physical Review E} \textbf{\bibinfo{volume}{81}},
  \bibinfo{pages}{041129} (\bibinfo{year}{2010}).

\bibitem[{\citenamefont{Flomenbom and Taloni}(2008)}]{flomenbom2008single}
\bibinfo{author}{\bibfnamefont{O.}~\bibnamefont{Flomenbom}} \bibnamefont{and}
  \bibinfo{author}{\bibfnamefont{A.}~\bibnamefont{Taloni}},
  \bibinfo{journal}{EPL (Europhysics Letters)} \textbf{\bibinfo{volume}{83}},
  \bibinfo{pages}{20004} (\bibinfo{year}{2008}).

\bibitem[{\citenamefont{Lizana et~al.}(2010)\citenamefont{Lizana,
  Ambj{\"o}rnsson, Taloni, Barkai, and Lomholt}}]{lizana2010foundation}
\bibinfo{author}{\bibfnamefont{L.}~\bibnamefont{Lizana}},
  \bibinfo{author}{\bibfnamefont{T.}~\bibnamefont{Ambj{\"o}rnsson}},
  \bibinfo{author}{\bibfnamefont{A.}~\bibnamefont{Taloni}},
  \bibinfo{author}{\bibfnamefont{E.}~\bibnamefont{Barkai}}, \bibnamefont{and}
  \bibinfo{author}{\bibfnamefont{M.}~\bibnamefont{Lomholt}},
  \bibinfo{journal}{Physical Review E} \textbf{\bibinfo{volume}{81}},
  \bibinfo{pages}{051118} (\bibinfo{year}{2010}).

\bibitem[{\citenamefont{Aslangul}(2007)}]{aslangul2007classical}
\bibinfo{author}{\bibfnamefont{C.}~\bibnamefont{Aslangul}},
  \bibinfo{journal}{EPL (Europhysics Letters)} \textbf{\bibinfo{volume}{44}},
  \bibinfo{pages}{284} (\bibinfo{year}{2007}).

\bibitem[{\citenamefont{Abramowitz and Stegun}(1965)}]{abramowitz1965handbook}
\bibinfo{author}{\bibfnamefont{M.}~\bibnamefont{Abramowitz}} \bibnamefont{and}
  \bibinfo{author}{\bibfnamefont{I.}~\bibnamefont{Stegun}},
  \emph{\bibinfo{title}{Handbook of mathematical functions: with formulas,
  graphs, and mathematical tables}}, vol.~\bibinfo{volume}{55}
  (\bibinfo{publisher}{Dover publications}, \bibinfo{year}{1965}).

\end{thebibliography}

\end{document}